\documentclass[useAMS,usenatbib]{mn2e}

\usepackage{epsfig,graphicx,amsmath,amssymb}
\usepackage{times}
\usepackage{natbib}

\def\JCAP{J. Cosmology  Astropart. Phys.}
\def\prd{Phys. Rev. D}

\def\apj{ApJ}

\def\apjl{ApJ}
\def\mnras{MNRAS}

\def\aj{AJ}

\def\physrep{Phys. Rep.}

\title[The effects of primordial non-Gaussianity on the cosmological
reionization] {The effects of primordial non-Gaussianity on the
  cosmological reionization} \author[D.~Crociani et al.]
{D. Crociani$^{1,2}$,
  L. Moscardini$^{1,2}$, M. Viel$^{3,4}$ and S. Matarrese$^{5,6}$\\
  $^1$ Dipartimento di Astronomia, Universit\`a di Bologna, via Ranzani 1, I-40127 Bologna, Italy (daniela.crociani5@unibo.it, lauro.moscardini@unibo.it) \\
  $^2$ INFN/National Institute for Nuclear Physics, Sezione di Bologna, viale Berti Pichat 6/2, I-40127 Bologna, Italy \\
  $^3$ INAF - Osservatorio Astronomico di Trieste, Via G.B. Tiepolo 11, I-34131 Trieste, Italy
  (viel@oats.inaf.it)\\
  $^4$ INFN/National Institute for Nuclear Physics, Sezione di Trieste, Via Valerio 2, I-34127 Trieste, Italy \\
  $^5$ Dipartimento di Fisica, Universit\`a di Padova, Via Marzolo 8, I-35131 Padova, Italy (sabino.matarrese@pd.infn.it) \\
  $^6$ INFN/National Institute for Nuclear Physics, Sezione di Padova, Via Marzolo 8, I-35131 Padova, Italy}
\begin{document}

\date{Accepted ???. Received ???; in original September 2008}

\pagerange{\pageref{firstpage}--\pageref{lastpage}} \pubyear{2008}

\maketitle

\label{firstpage}

\begin{abstract}

  We investigate the effects of non-Gaussianity in the
  primordial density field on the reionization history. We rely on a
  semi-analytic method to describe the processes acting on the
  intergalactic medium (IGM), relating the distribution of the
  ionizing sources to that of dark matter haloes.  Extending previous
  work in the literature, we consider models in which the primordial
  non-Gaussianity is measured by the dimensionless non-linearity
  parameter $f_{\rm NL}$, using the constraints recently obtained from
  cosmic microwave background data.  We predict the ionized fraction
  and the optical depth at different cosmological epochs assuming two
  different kinds of non-Gaussianity, characterized by a
  scale-independent and a scale-dependent $f_{\rm NL}$ and comparing
  the results to those for the standard Gaussian scenario.  We
  find that a positive $f_{\rm NL}$ enhances the formation of
  high-mass haloes at early epochs, when reionization
  begins, and, as a consequence, the IGM ionized fraction can grow by
  a factor up to 5 with respect to the corresponding Gaussian model.
  The increase of the filling factor has a small impact on the
  reionization optical depth and is of order $\sim 10$ per cent
  if a scale-dependent non-Gaussianity is assumed.  Our predictions
  for non-Gaussian models are in agreement with the latest WMAP results
  within the error bars, but a higher precision is required to
  constrain the scale dependence of non-Gaussianity.

\end{abstract}


%%%%%%%%%%%%%%%%%%%%%%%%%%%%%%%%%%%%%%%%%%%%%%%%%%%%%%%%%%%%%%%%%%%
%%%%%%%%%%%%%%%%%%%% Introduction %%%%%%%%%%%%%%%%%%%%%%%%%%%%%%%%%
%%%%%%%%%%%%%%%%%%%%%%%%%%%%%%%%%%%%%%%%%%%%%%%%%%%%%%%%%%%%%%%%%%%

\section{Introduction} \label{sect:intro}

Reionization marks a crucial event in the history of the universe,
when the first sources of ultra-violet (UV) radiation ionize the
neutral Intergalactic Medium (IGM) and affect the subsequent formation
of the cosmic structures. When reionization ends,
the small amount of left neutral hydrogen is
responsible for the absorption lines that we observe today in the
spectra of far objects. However, the way in which this complex
phenomenon occurs is still not well understood and
the most recent observations paint it as a spatially
inhomogeneous and not istantaneous process. While the Gunn \& Peterson
trough of the high-$z$ QSO spectra suggests a late epoch of
reionization at $z\approx 6$
\citep{fan2001,becker2001,white2003b,fan2006}, the very recent
analysis of the 5-year WMAP data on the cosmic microwave background
(CMB) polarization shows an IGM optical depth $\tau\sim 0.084$ which
is in better agreement with an earlier reionization redshift, $z\sim
10.8$ \citep{komatsu2008}.  On the other hand, a late reionization end
at $z\sim 6$ is also probed by the IGM temperature measured at $z<4$
\citep{hui2003} and by the lack of evolution in the luminosity
function of Lyman-$\alpha$ galaxies between $z=5.7$ and $z=6.5$
\citep[][see, however, {\protect \cite{ota2008}} for evidences of a
  decline at high $z$]{malhotra2004}. Overall, the present situation
regarding reionization at redshift $z\sim6$ as probed by QSO spectra
is still unclear \citep{becker2006}.

Many analytic, semi-analytic and numerical models \citep[see, e.g.][]{
  gnedin2000,ciardi2003b,wyithe2003,barkana2004,
  haiman2003,madau2004,Wyithe2006,choudhury2007,iliev2007,ricotti2008}
have been proposed to describe this poorly understood reionization
process. They basically relate the statistical properties and
morphology of the ionized regions to the hierarchical growth of the
ionizing sources, making more or less detailed assumptions to describe
the ionization and recombination processes acting on the IGM.  Since
the first sources of UV background radiation appear in the firstly
formed dark matter haloes, which correspond to the highest peaks of
the primordial density field, the reionization process is expected to
strongly depend on the main parameters describing the cosmological
model and the power spectrum of primordial density fluctuations.  For
instance, the possible presence of an evolving component of dark
energy can imprint signatures in the resulting morphology of the
ionized regions and change the time-scales of the whole process
\citep[see, e.g.][]{maio2006,crociani2008}. Also the nature and
  the statistical distribution of the primordial matter fluctuations
  can influence the reionization history. Although the standard
  scenario for the origin of the structures assumes that the
  primordial perturbations are adiabatic and have a (almost) Gaussian
  distribution, small deviations from primordial Gaussianity affect
  the dark halo counts, in the rare-event tail, thus also in the high
  peaks of the density field which originated collapsed objects at
  high $z$.

Aim of this work is to investigate the effects of some level of non-Gaussianity in 
the primordial density field on the reionization history. We will make use of analytical
techniques to describe the processes in action on the IGM. In particular, 
we will extend previous works in which the considered
non-Gaussian models have density fluctuations described by a renormalized 
$\chi^{2}$ probability distribution with $\nu$
degrees of freedom \citep{avelino2006} or by a modified Poisson distribution 
with a given expectation value $\lambda$ \citep{chen2003}.
Here we will adopt a more convenient
way to introduce primordial non-Gaussianity, which has now become standard in
the literature, based on the parameter $f_{\rm NL}$ (see the next section
for its definition). In particular, we will assume two different kinds
of non-Gaussianity, characterized by a
scale-independent and a scale-dependent $f_{\rm NL}$ parameter.

The paper is organised as follows. In Section \ref{sect:ng} we
introduce the main characteristics of the cosmological models with
primordial non-Gaussianity considered here. Section \ref{sect:model}
reviews the main assumptions of the analytical model adopted to
describe the cosmic reionization process. The main results of 
our analysis are presented and discussed in Section \ref{sect:res}. 
Finally in Section \ref{sect:conclu} we draw our conclusions.

%%%%%%%%%%%%%%%%%%%%%%%%%%%%%%%%%%%%%%%%%%%%%%%%%%%%%%%%%%%%%%%%%%%
%%%%%%%%%%%%%%%%  non-Gaussianity %%%%%%%%%%%%%%%%%%%%%%%%%%%%%%%%%
%%%%%%%%%%%%%%%%%%%%%%%%%%%%%%%%%%%%%%%%%%%%%%%%%%%%%%%%%%%%%%%%%%%

\section{Modeling primordial non-Gaussianity} \label{sect:ng}

The main purpose of this work is to study the process of reionization
under the assumption that the formation of the first ionizing sources
is driven by the spherical collapse of overdense regions 
in a non-Gaussian primordial density field. 
The predicted reionization history will be compared to
that obtained assuming the `standard' model, with a Gaussian
distribution of primordial perturbations, that
will represent our `reference' case.

All the models considered in this work share the cosmological
parameters suggested by the recent analysis of the 5-year WMAP data
\citep{komatsu2008}: a $\Lambda$CDM cosmology where the contributions
to the present density parameter from dark matter, cosmological
constant and baryons are $\Omega_{m0}=0.279$,
$\Omega_{\Lambda0}=0.7214$, $\Omega_{b0}=0.0461$, respectively; the
Hubble constant (in units of 100 km/s/Mpc) is $h=0.701$.  The
normalization of the cold dark matter power spectrum is fixed by
assuming $\sigma_{8}=0.817$ and the primordial spectral index is taken
to be $n=0.96$.

We will describe the level of primordial
non-Gaussianity using the dimensionless parameter $f_{\rm NL}$ which
weighs the quadratic correction to the linear Gaussian term in Bardeen's gauge-invariant
potential $\Phi$:
\begin{equation}\label{eq:1a}
  \Phi(\mathbf{x})=\Phi_{G}(\mathbf{x})+
  f_{\rm NL}*(\Phi^2_{G}(\mathbf{x})-\langle\Phi^2_{G}(\mathbf{x})\rangle)\ ,
\end{equation}
where $\Phi_{G}(\mathbf{x})$ is a Gaussian random field and $*$
denotes a convolution. On scales smaller than the
Hubble radius, $\Phi$ is minus the usual Newtonian
gravitational potential.  With our convention, a positive value for
$f_{\rm NL}$ leads to a positive skewness for the distribution of the
matter density fluctuations. 

As shown by eq.(\ref{eq:1a}), in general the non-Gaussian
contribution to the gravitational potential $\Phi$ can be written as a
convolution between a space- and/or shape-dependent $f_{\rm 
NL}(\mathbf{x})$ and the quadratic term
$\Phi_{G}^2(\mathbf{x})$. The possible dependences of $f_{\rm
  NL}$ are often neglected in the literature: this is done mostly
for sake of simplificity, but it can motivated by the small
{\it r.m.s.} value of $\Phi$. In this case the bispectrum of the
gravitational potential, defined as
\begin{equation}\label{eq:1e}
  \langle \Phi({\mathbf k_{1}})\,\Phi({\mathbf k_{2}})\,\Phi({\mathbf k_{3}}) 
  \rangle=
  (2\pi)^3\delta^3({\mathbf k_{1}},{\mathbf k_{2}},
  {\mathbf k_{3}})F_{s}(k_{1},k_{2},k_{3})\ ,
\end{equation}
$\delta^3$ being Dirac's delta function, assumes a
dependence on the wavenumbers called {\it local shape}, for which the
term $F_{s}(k_{1},k_{2},k_{3})$ can be expressed as
\begin{equation}\label{eq:1f}
  F_{s}(k_{1},k_{2},k_{3})=2f_{\rm NL}
  [P(k_{1})P(k_{2})+P(k_{1})P(k_{3})+P(k_{2})P(k_{3})]\ .
\end{equation}
In the previous equation $P(k)\equiv \Delta_{\Phi}k^{-3+(n-1)}$
represents the normalized power-spectrum of $\Phi$. Bispectra which
can be expressed like in eq.(\ref{eq:1f}) are typical for models where
the non-Gaussianity is produced outside the horizon or when the
inflaton has a varying decay rate. It can be shown that
eq.(\ref{eq:1f}) assumes the largest values for squeezed
configurations, i.e. when one wavenumber is much smaller  
than the other two.

Alternative models for primordial non-Gaussianity, based on a single
field with higher derivative terms, predict a different shape for the
bispectrum, having the so-called {\it equilater shape}.  Its expression
can be still obtained using eq.(\ref{eq:1e}), but replacing
$F_{s}(k_{1},k_{2},k_{3})$ as follows \citep{creminelli2007}:
\setlength\arraycolsep{1pt}
\begin{eqnarray}\label{eq:1g}
  F_{s}(k_{1},k_{2},k_{3})& = & 6f_{\rm NL}(k_{1},k_{2},k_{3})\Delta^2_{\Phi}\Bigg[-\frac{(k_{1}k_{2})^{n-1}}{(k_{1}k_{2})^{3}}+2\ \mathrm{perm}-\nonumber\\
  &-&\frac{2(k_{1}k_{2}k_{3})^{2(n-1)/3}}{(k_{1}k_{2}k_{3})^{2}}
  +\frac{(k_{1}^{1/3}k_{2}^{2/3}k_{3})^{(n-1)}}{k_{1}k_{2}^{2}k_{3}^{3}}+\nonumber\\
  &+&5\ \mathrm{perm}\Bigg]\ ,
\end{eqnarray}
where we explicitly write the possible scale-dependence of $f_{\rm
  NL}$.  As its name suggests, in this case the maximum amplitude of
the bispectrum is where the wavenumbers are all equal.  We refer to 
\cite{loverde2008} for an extended discussion about the role of
bispectrum shapes in the parametrization of primordial non-Gaussianity
and \cite{bartolo2004} for a review on the predictions 
for $f_{\rm NL}$ in different inflationary models.

At present, the strongest constraints on the parameter $f_{\rm NL}$
are based on CMB data. Analysing the 5-year temperature maps obtained
by the satellite WMAP, \cite{komatsu2008} derived $-9<f_{\rm NL}<111$
when the local shape for the bispectrum is assumed, and $-151<f_{\rm
  NL}<253$ for the equilateral shape. Both limits have been estimated
at the 95 per cent confidence level.  The fact that the presence of
some amount of primordial non-Gaussianity alters the growth of density
fluctuations and then the formation and evolution of cosmic
structures, suggests that the large-scale structure (LSS) of the
universe can be an alternative powerful probe for $f_{\rm NL}$, which
has also the important advantage of being based mostly on
three-dimensional data. Many theoretical studies, based both on
analytic and numerical analyses, have investigated the constraining
capability of different observables like the abundances of virialised
objects like clusters
\citep{messina1990,moscardini1991,weinberg1992,matarrese2000,
  verde2000, mathis2004, kang2007, grossi2007, loverde2008}, halo
biasing \citep{dalal2008,matarrese2008,mcdonald2008}, galaxy
bispectrum \citep{sefusatti2007}, density mass field distribution
\citep{grossi2008} and topology \citep{matsubara2003,hikage2008},
integrated Sachs-Wolfe effect \citep{afshordi2008, carbone2008}, low
density intergalactic medium and the Lyman-$\alpha$ flux
\citep{viel2008}, 21-centimeter fluctuations
\citep{cooray2005,pillepich2007}, and reionization, as discussed in
this paper. In general, the application of these theoretical results
to real LSS data provided weaker constraints on $f_{\rm NL}$ with
respect to the CMB. The only exception is the very recent analysis
made by \cite{slosar2008}, who applied the bias formalism to a
compendium of large-scale data, including the spectroscopic and
photometric luminous red galaxies from the Sloan Digital Sky Survey
(SDSS), the SDSS photometric quasars and the cross-correlation between
galaxies and dark matter via Integrated Sachs-Wolfe
effect. Considering the local shape only, they found $-29<f_{\rm
  NL}<70$ (at 95 per cent confidence level).  It is important to
notice that the scales probed by CMB and LSS are generally different
and can give complementary information on $f_{\rm NL}$ if the
primordial non-Gaussianity is assumed to be scale-dependent \citep[see
the discussion in][]{loverde2008}.

\begin{figure*}
\includegraphics[height=7cm, width=8cm]{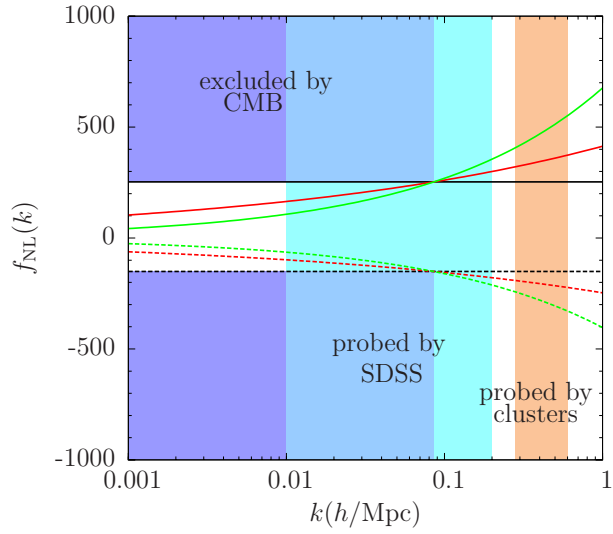}
\caption{ The scale dependence of the non-Gaussianity parameter
  $f_{\rm NL}$ for the models with equilateral shape considered in
  this paper.  Results for two different choices of the pivoting
  value, $f_{\rm NL}=-151$ and $f_{\rm NL}=253$, are shown by dashed
  and solid curves, respectively. Different values for the slope
  $\alpha$ have also been used: $\alpha=0$ (black lines),
  $\alpha=-0.1$ (red lines) and $\alpha=-0.2$ (green lines).  The
  shaded regions on the right show the scales probed by the SDSS
  (cyan) and the galaxy clusters (orange), while the blue region
  refers to the range excluded at 95 per cent confidence level by the
  CMB data \citep{komatsu2008}.}
\label{fig:00a}
\end{figure*}

In this work, we consider non-Gaussian model with bispectrum having
both the local and equilateral shapes. We will use values for $f_{\rm
  NL}$ in the range constrained by the 5-year WMAP results
\citep{komatsu2008}, i.e. $-9<f_{\rm NL}<111$ and $-151<f_{\rm
  NL}<253$ for local and equilateral shapes, respectively. In the last
case, we also allow the non-Gaussianity to vary with the scale,
assuming the dependence proposed by \cite{loverde2008}, namely:
\begin{equation}\label{eq:1h}
  f_{\rm NL}(k_{1},k_{2},k_{3})=f_{\rm NL}
  \Bigg(\frac{k_{1}+k_{2}+k_{3}}{k_{\rm CMB}}\Bigg)^{-2\alpha}\ .
\end{equation}
The normalisation of the previous relation is chosen in order to avoid
violating the WMAP constraints: for this reason $f_{\rm NL}$
represents the equilateral parameter measured on the $k_{\rm CMB}$
scale of $0.086 h/\,$Mpc, roughly corresponding to largest multipole used by
\cite{komatsu2008} to estimate the non-Gaussianity in the WMAP data,
$\ell=700$.  The slope $\alpha$ is a free parameter, assumed to be
constant, such that $| \alpha | \ll 1$ between CMB and cluster scales.
Following \cite{loverde2008} we consider small negative values for
$\alpha$, to enhance the non-Gaussianity on scales smaller
than CMB.  The resulting behaviour for the $f_{\rm NL}$ parameter is
shown in Fig.\ref{fig:00a}, where we assume $\alpha=0, -0.1, -0.2$,
for the slope of the scale dependence, and $f_{\rm NL}=-151$, and
$f_{\rm NL}=253$ as pivoting values at the CMB scale, in agreement
with the WMAP equilateral constraints.  It is evident from the plot
that, with our assumption for the scale-dependence relation of
non-Gaussianity, the absolute value of $f_{\rm NL}$ at the scales
relevant for the halo formation, and then for reionization, can be a
factor 2-3 larger than the maximum amount directly derived from the
CMB analysis: this can amplify the possible effects of primordial
non-Gaussianity. Furthermore, we note that the possible
non-Gaussianity probes also are based on observational data coming
from different ranges of redshift.

Deviations from Gaussianity influence the evolution of the density
fluctuations, and affect the distribution of the virialized dark
matter haloes at a given cosmological epoch. This reflects on the mass
function: its high-mass tail is enhanced (reduced) in case of positive
(negative) values of $f_{\rm NL}$. While in the Gaussian case,
  the \cite{press1974} (PS74 hereafter) approach, together with its
  modern improvements \citep{lacey1993,
    sheth1999,jenkins2001,warren2006}, represent powerful tools to
  model the evolution of the ionizing sources, for mildly non-Gaussian
  fields it is possible to make use of the analytic relation found by
  \cite{matarrese2000} extending the PS74 formalism, which has been
positively tested against the results of high-resolution N-body
simulations by \cite{grossi2007}. If $n_G(M,z)$ represents the density
of dark matter haloes with mass $M$ at redshift $z$ as obtained
assuming Gaussian initial conditions, here modeled assuming the
relation found by \cite{sheth1999}, the corresponding expression for a
non-Gaussian models having the same cosmological parameters can be
written as
\begin{equation}\label{eq:1c}
n_{\rm NG}(M,z)=F_{\rm NG}(M,z)n_{\rm G}(M,z)\ ,
\end{equation}
where the correction factor $F_{\rm NG}$ is given by
\begin{equation}\label{eq:1d}
  F_{\rm NG}(M,z) \simeq \Bigg | \Bigg[\frac{1}{6}\frac{\delta^2_{c}}{\delta_{*}}\frac{d S_{3}(M)}{d \ln\sigma_{M}}
  +\frac{\delta_{*}}{\delta_{c}}\Bigg]\Bigg |\exp\Bigg(\frac{\delta^{3}_{c}S_{3}}{6\sigma^{2}_{M}}\Bigg)\ .
\end{equation}
In the previous relation
$\delta_{*}\equiv \delta_{c}\sqrt{1-S_{3}(M)\delta_{c}/3}$, $\delta_{c}$
represents the collapse threshold at $z$, $\sigma_{M}^2$ is the mass
variance at $z=0$ and $S_{3}(M)$ is the normalized skewness of the
primordial density field on mass scale $M$, namely $S_{3}(M)=-f_{\rm
  NL}\mu_{3}(M)/\sigma^4_{M}$.  Then in order to compute the mass function for
non-Gaussian models it is necessary to evaluate the third-order moment
$\mu_{3}$, that depends on the bispectrum of the gravitational
potential
$\langle\Phi(\mathbf{k_{1}})\Phi(\mathbf{k_{2}})\Phi(\mathbf{k_{3}})\rangle$:
\begin{eqnarray}\label{eq:1b}
  \mu_{3}(M) & = & \int \frac{{\mathrm d}\mathbf{k_{1}}}{(2\pi)^3} 
  \int \frac{{\mathrm d}\mathbf{k_{2}}}{(2\pi)^3} 
  \int \frac{{\mathrm d}\mathbf{k_{3}}}{(2\pi)^3}\times \nonumber\\
  & & W(k_{1})W(k_{2})W(k_{3})F(k_{1})F(k_{2})F(k_{3})\times \nonumber\\
  & & \langle\Phi(\mathbf{k_{1}})\Phi(\mathbf{k_{2}})\Phi(\mathbf{k_{3}})\rangle
\end{eqnarray}
where $W(k)$ is the Fourier transform of a spherical top-hat function
on the mass scale $M$, $F(k)\equiv T(k)g(k)$, being $T(k)$ the cold
dark matter transfer function and $g(k)\equiv
-2(k/H_{0})^{2}/(3\Omega_{m0})$ is required to go from the gravitational
potential to the density via the Poisson equation.

\begin{figure*}
\includegraphics[height=12cm, width=9cm]{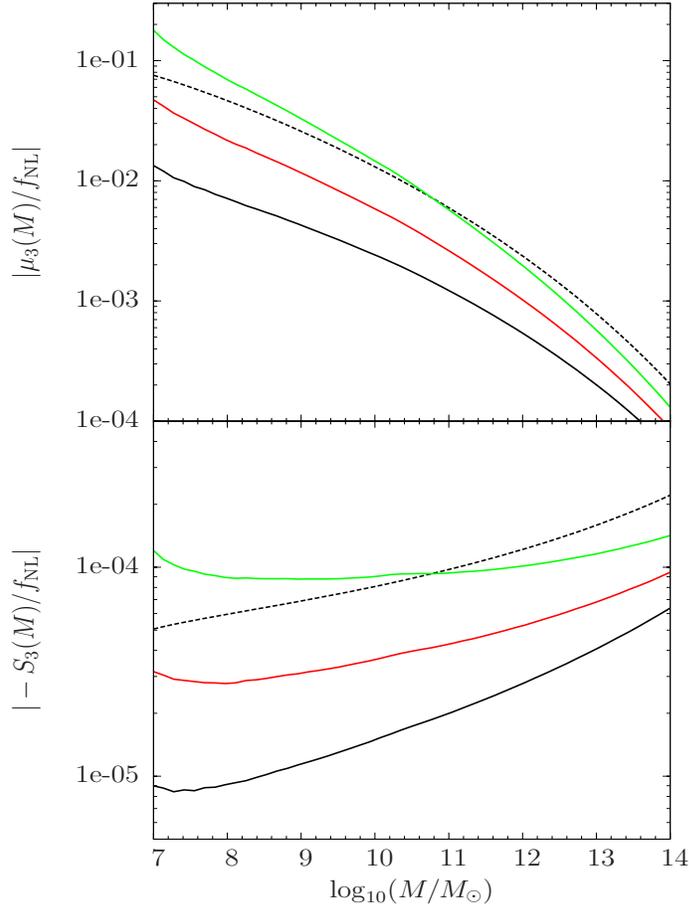}
\caption{The skewness $\mu_3$ (top panel) and the normalized skewness
  $S_3$ (bottom panel), given per unit $f_{\rm NL}$. The black dashed
  curve refers to the model with local shape, while the coloured solid
  curves present the results for the equilateral configuration with
  $\alpha=0, -0.1, -0.2$ (black, red and green lines,
  respectively).}\label{fig:0a}
\end{figure*}

In Fig.\ref{fig:0a} we show, as a function of the halo mass, $\mu_{3}$
and $S_{3}$, for models with both local and equilateral shapes. Both
skewness parameters are given per unit non-Gaussianity parameter
$f_{\rm NL}$; for the equilateral case we also consider the
possibility of scale-dependence for the non-Gaussian term.  As already
shown by \cite{loverde2008}, the two classes of models give quite
different predictions for both the amplitude and the mass dependence
of the two considered quantities, but this discrepancy decreases as
the mass scale increases, since the local and the equilateral cases
become more and more similar.  The scale-dependence of $f_{\rm NL}$
strongly affects the non-Gaussianity contribution for the smaller
masses scales and this effect grows when higher negative $\alpha$
parameters are assumed.

\begin{figure*}
\includegraphics[height=9cm, width=15cm]{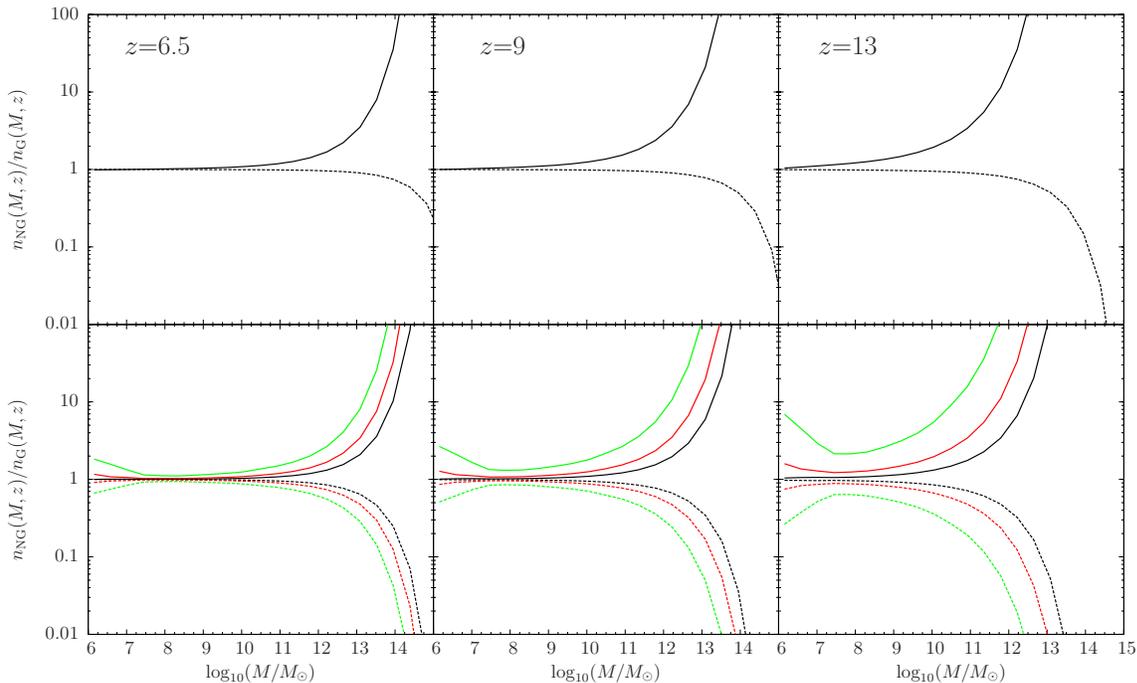}
\caption{The ratio between the dark matter halo mass functions for
  non-Gaussian and Gaussian models, computed at $z=6.5$ (left panels),
  $z=9$ (central panels) and $z=13$ (right panels).  Top panels show
  the results for non-Gaussian models with local shape, where $f_{\rm
    NL}=-9$ (dashed lines) and $f_{\rm NL}=111$ (solid lines) are
  assumed. In the bottom panels, which refer to non-Gaussian models
  with equilateral shape, dashed and solid lines correspond to $f_{\rm
    NL}=-151,253$, respectively; black, red and green lines refer to
  $\alpha=0$ (i.e. no scale-dependence), $\alpha=-0.1$ and
  $\alpha=-0.2$ respectively.}

\label{fig:1a}
\end{figure*}

Inserting the values for $S_3$ in eq.(\ref{eq:1d}), we can estimate
the effects of primordial non-Gaussianity on the dark matter halo
distribution at the cosmological epochs relevant for the process of
reionization.  The results, shown in terms of ratio with respect to
the Gaussian predictions, are shown in Fig.\ref{fig:1a}, for both
local and equilateral shapes (upper and lower panels, respectively).
Here we adopt for the $f_{\rm NL}$ parameter at the CMB scale the
values corresponding to the 95 per cent confidence level, as derived
from the 5-year WMAP data:$f_{\rm NL}=-9, 111$ for the local shape,
and $f_{\rm NL}=-151, 253$ for equilateral one.  Since the mass
density probability function is positively skewed in case of positive
($f_{\rm NL}>0$) non-Gaussian contributions, the probability of
overcoming the collapse threshold becomes higher.  As a consequence,
the formation of high-mass haloes is enhanced and anticipated when
$f_{\rm NL}>0$. Fig. \ref{fig:1a} shows that the mass function can be
increased by a factor of 10 at $z=13$ for haloes with mass $M\sim
10^{11}M_{\odot}$ when compared to the standard scenario.  We should
however remark that high-mass haloes ($M>10^{9}M_{\odot}$) at early
cosmological epochs are rare events, as shown also by the small number
density at $z=13$ in the reference case, $n(>10^{9}M_{\odot})\lesssim
5 \times 10^{-3}$/Mpc$^{3}$.  Then this effect is expected to have a little
impact on integrated quantities as the total ionized fraction ot the
IGM optical depth. Unlike the local model, the scale dependence of
non-Gaussianity increases the abundance of the low-mass haloes by a
factor of $\sim 10$ at $z=13$ when compared to the standard case.  The
opposite applies for $f_{\rm NL}<0$.  As already noticed by
\cite{matarrese2000} [see also \cite{verde2001,grossi2007}], this
effect is more evident at early cosmological epochs, exactly when the
process of IGM ionization starts. For this reason a non-Gaussian
distribution of the primordial density field can affect the way in
which reionization occurs, leaving its imprints on it, as we will
investigate in the next sections.

%%%%%%%%%%%%%%%%%%%%%%%%%%%%%%%%%%%%%%%%%%%%%%%%%%%%%%%%%%%%%%%%%%%
%%%%%%%%%%%%%%%%%%%%% AL06 %%%%%%%%%%%%%%%%%%%%%%%%%%%%%%%%%%%%%%%%
%%%%%%%%%%%%%%%%%%%%%%%%%%%%%%%%%%%%%%%%%%%%%%%%%%%%%%%%%%%%%%%%%%%

\section{An analytic approach to cosmic reionization}
\label{sect:model}

In this section, we briefly review the main assumptions underlying the
analytic model adopted to describe the process of cosmic
reionization. This model is based on the approach proposed by
\cite{avelino2006} [see also \cite{haiman2003,chen2003} for further
details]; our implementation, however, differs in some
aspects which will be discussed later.

In this model, the statistical properties of the ionized regions are
related to the hierarchical growth of the ionizing sources through
simple assumptions on how the galaxies ionize the IGM and on how the
IGM recombines.  A one-to-one correspondence between the distribution
of galaxies and HII regions is established, such that a single galaxy
of mass $M_{\rm gal}$ can ionize a region of mass $M_{\rm HI\!I}=\zeta
M_{\rm gal}$. Here $\zeta$ represents the ionization efficiency of the
galaxy, and it is strictly dependent on the nature of the ionizing
sources. We will take it as a constant, fixed in such a way that
reionization ends at $z=6.5$.

Since at high $z$ the cooling of the gas becomes efficient in haloes
having a virial temperature $T\ge 10^4$ K, unlike in
\cite{avelino2006}, in our analysis we consider only type Ia ($10^4$
K $\le T \le 9 \times10^4$ K) and type Ib ($T > 9\times 10^4$ K) haloes,
neglecting the contribution of the type II sources, which would
correspond to haloes with $400$ K $\le T \le 10^4$ K. We recall
that the distinction between the halo types is related to the way in
which they impact the IGM: type Ia sources can grow only in neutral
regions, while type Ib haloes can appear also in ionized
regions. Consequently they affect differently the ionization phases of
IGM.

The total collapsed fraction $F_{\rm coll}(z)$ at different redshifts
can be computed using eq.(\ref{eq:1c}):
\begin{eqnarray}\label{eq:2a}
  F_{\rm coll}(z)&=&\frac{1}{\bar{\rho}_{0}}\int_{M_{\rm min}(z)}^{\infty}
  \mathrm{d}M \, n_{\rm NG}(M,z) \nonumber\\
  &=&\frac{1}{\bar{\rho}_{0}}\int_{M_{\rm min}(z)}^{\infty}
  \mathrm{d}M \, n_{\rm G}(M,z)F_{\rm NG}(M,z)\ ,
\end{eqnarray}
where $\bar{\rho}_{0}$ is the present-day matter density and $M_{\rm
  min}(z)$ is the minimum mass corresponding to the virial temperature
$T$, which can be computed by inverting the relation proposed by
\cite{barkana2001}, namely:
\begin{equation}\label{eq:3a}
T=1.98 \times 10^4 \left( \frac{1+z}{10}\right) \Bigg(\frac{M}{10^8 M_{\odot}h^{-1}}\Bigg)^{2/3}
\Bigg(\frac{\Omega_{\rm m0}}{\Omega_{\rm m}^{z}}\frac{\Delta_{c}}{18\pi^{2}}\Bigg)^{1/3}
\mathrm{K}\ .
\end{equation}
In the previous equation, $\Delta_{\rm c}$ represents the virial overdensity at redshift $z$
and $\Omega_{\rm m}^{z}$ is the matter density parameter at redshift $z$.

Consequently, the collapsed fractions in Ia and Ib haloes are given
by
\begin{eqnarray}\label{eq:2b}
F_{{\rm coll},Ib}(z)&=&\frac{1}{\bar{\rho}_{0}}\int_{M_{{\rm min},Ib}(z)}^{\infty}
\mathrm{d}M \, n_{\rm NG}(M,z)\nonumber\\
F_{{\rm coll},Ia}(z)&=&\frac{1}{\bar{\rho}_{0}}\int_{M_{{\rm min},Ia}(z)}^{\infty}
\mathrm{d}M \, n_{\rm NG}(M,z)-F_{{\rm coll},Ib}(z)\ , \nonumber\\
\end{eqnarray}
where $M_{{\rm min},Ib}$ and $M_{{\rm min},Ia}$ are the minimum masses
for Ib and Ia sources, obtained using in eq.(\ref{eq:3a}) $T= 9\times
10^4$ and $10^4$ K, respectively.

The action of the ionizing sources is smoothed down by the
recombination of the IGM, here considered as a homogeneous gas.  The
recombination rate is linearly dependent on the IGM clumping factor
$C_{\rm HI\!I}=<n_{\rm HI\!I}^2>/<n_{\rm HI\!I}>^2$, for which, following \cite{haiman2006}, we assume a
redshift evolution modeled as:
\begin{equation}\label{eq:2c}
C_{\rm HI \!I}(z)=1+9 \Bigg(\frac{1+z}{7}\Bigg)^{-\beta}\ ,
\end{equation} 
being $\beta$ a free parameter.  As shown by \cite{avelino2006}, the
predicted reionization history of the universe has significant
uncertainties introduced by the poor knowledge of the $z$-dependence
of the clumping factor, which cannot be robustly constrained even
considering the 3-year WMAP results for the reionization optical depth. Since
they found good consistency between predicted and observed optical
depths irrespectively of the amount of primordial non-Gaussianity in
the models they consider, we decide to set $\beta=0$. In this case,
the $z$-dependence of $C_{\rm HI\!I}$ is neglected and $C_{\rm HI\!I}=10$.
The impact of the assumption of a constant clumping factor will be
discussed later.

The probability that a photon emitted
at a given cosmological epoch $z_{\rm i}(t_{\rm i})$ is still ionizing at
$z<z_{\rm i}$ can be written as
\begin{equation}\label{eq:2d}
P(t_{\rm i},t)=\mathrm{exp}\Bigg(\frac{t_{\rm r}}{t}-\frac{t_{\rm r}}{t_{\rm i}}\Bigg)\ ,
\end{equation}
with $t_{\rm r}=\alpha_{\rm B} C_{\rm HI\!I} n_{\rm HI}(t_{\rm i})t_{\rm i}^{2}$, being
$\alpha_{\rm B}$ the recombination coefficient of HI ($=2.6 \times
10^{-13}$ cm$^{3}$/s at $T=10^4$ K) and
$n_{\rm HI}(z)=1.88\Omega_{\rm b0}h^{2}/0.022(1+z)^{3}$/cm$^3$ the hydrogen
density at redshift $z$.  Then, the filling factor $F_{\rm HI\!I}(z)$ at a
given cosmological epoch is
\begin{eqnarray}\label{eq:2e}
F_{\rm HI\!I}(z)&=&\int_{\infty}^{z}\mathrm{d}z' \zeta \, 
\Bigg \{\frac{\mathrm{d}F_{{\rm coll},Ib}}{\mathrm{d}z'}(z')+
[1-F_{\rm HI\!I}(z')]\times\nonumber\\
&\times&\frac{\mathrm{d}F_{{\rm coll},Ia}}{\mathrm{d}z'}(z')\Bigg\}P(z',z)\ ,
\end{eqnarray}
where the ionizing efficiency is assumed to be the same for the
different types of haloes.  We notice that the different nature of
type Ib and Ia sources appears in the right-hand side of
eq.(\ref{eq:2e}), where the $(1-F_{\rm HI\!I})$ factor explicitly
considers that type Ia haloes form only in neutral regions.

Finally, the HII filling factor allows us to estimate the reionization optical
depth as follows:
\begin{eqnarray}\label{eq:2f}
  \tau(z)&=&c\sigma_{\rm T}\int_{\rm t}^{t_{0}}\mathrm{d}t' \,n_{\rm e}(t')\nonumber\\
  &=&1.08\,c\,\sigma_{\rm T}\int_{\rm z}^{0}\frac{\mathrm{d}t}{\mathrm{d}z'}\mathrm{d}z'
  \Bigg(1-\frac{3}{4}Y\Bigg)\frac{\rho_{\rm b}(z')}{m_{\rm p}}F_{\rm HI\!I}(z')\ ;
\end{eqnarray}
here $\sigma_{\rm T}$ represents the cross-section of the Thompson
scattering, $n_{\rm e}$ is the free-electron density, $c$ is the
speed of light, $Y$ is the Helium mass fraction, $\rho_{\rm b}(z)$ is
the baryon density at redshift $z$ and the factor 1.08 approximately
accounts for the contribution of the HeI reionization, assuming that
the HII and HeII fractions are equal and neglecting the effects of the
HeII to HeIII phase transition.

%%%%%%%%%%%%%%%%%%%%%%%%%%%%%%%%%%%%%%%%%%%%%%%%%%%%%%%%%%%%%%%%%%%
%%%%%%%%%%%%%%%%%%%%%%%%%%%%%% results %%%%%%%%%%%%%%%%%%%%%%%%%%%%
%%%%%%%%%%%%%%%%%%%%%%%%%%%%%%%%%%%%%%%%%%%%%%%%%%%%%%%%%%%%%%%%%%%

\section{Results and discussion}  \label{sect:res}

In this section we present and discuss the main results of the
application of the previous model under the assumption that the
structure formation history is starting from a primordial non-Gaussian
density field. The corresponding predictions will be compared to
the reionization scenario obtained for the Gaussian `reference' case.

\begin{figure*}
\includegraphics[height=10cm, width=8cm]{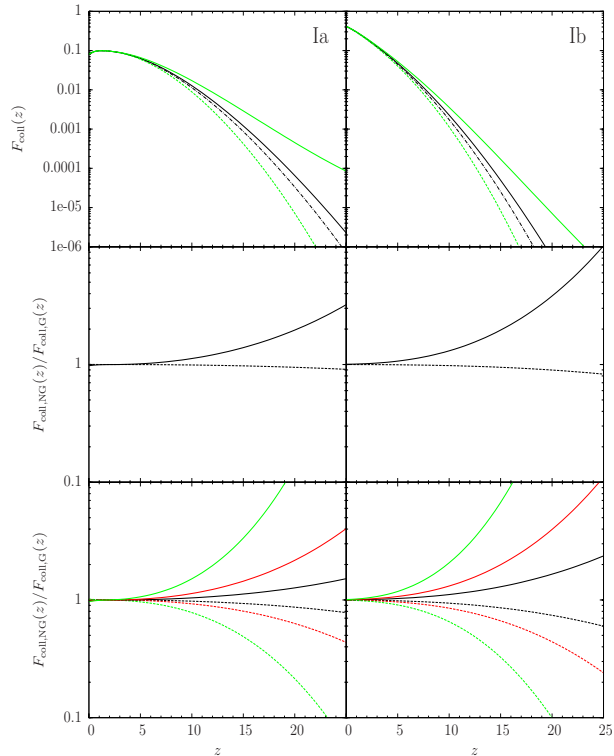}
\caption{
The collapsed fraction of Ia and Ib ionizing sources. The top panels
show the resulting $F_{\rm coll}$ for the Gaussian reference case
(dot-dashed line) and two extreme non-Gaussian models here
considered, which adopt $f_{\rm NL}=111$ with the local shape (solid
line) and $f_{\rm NL}=-151,253$ with $\alpha=-0.2$ for the equilateral
shape (green dotted and  solid surves, respectively). The other
panels show the ratio between the collapsed fractions of Ia (left
panels) and Ib (right panels) ionizing sources for non-Gaussian and
Gaussian models as a function of redshift.  In the middle panels the
results for non-Gaussian models with local shape are shown, where
$f_{\rm NL}=-9$ (dashed lines) and $f_{\rm NL}=111$ (solid lines) are
assumed. In the bottom panels, which refer to non-Gaussian models with
equilateral shape, dashed and solid lines correspond to $f_{\rm
NL}=-151,253$, respectively; black, red and green lines refer to
$\alpha=0$ (i.e. no scale-dependence), $\alpha=-0.1$ and $\alpha=-0.2$,
respectively.}
\label{fig:3a}
\end{figure*}

First of all, using the minimum collapsed mass at each cosmological
epoch as derived from the $M$-$T$ relation proposed by
\cite{barkana2001}, we can compute the total collapsed fractions for
different kinds of sources, following eq.(\ref{eq:2b}).  The results
are presented in Fig.\ref{fig:3a}, where we show the collapsed
fractions of Ia and Ib haloes for the local and the equilateral cases
here considered and their redshift evolution compared to the
predictions for the Gaussian scenario.  We can notice that for both
the local and the equilateral models, positive values for $f_{\rm NL}$
produce an enhancement of the collapsed fraction at a fixed redshift,
given the larger probability for high-mass haloes in the primordial
distribution.  When scale-dependent non-Gaussianity is considered,
this effect becomes higher as the power-law parameter $\alpha$
decreases. The opposite trend is observed when negative $f_{\rm NL}$
values are assumed: in this case we obtain a reduction of the
non-Gaussian collapsed fraction, due to the smaller probability for
high-mass haloes at high $z$.

\begin{figure*}
\includegraphics[height=5cm, width=17cm]{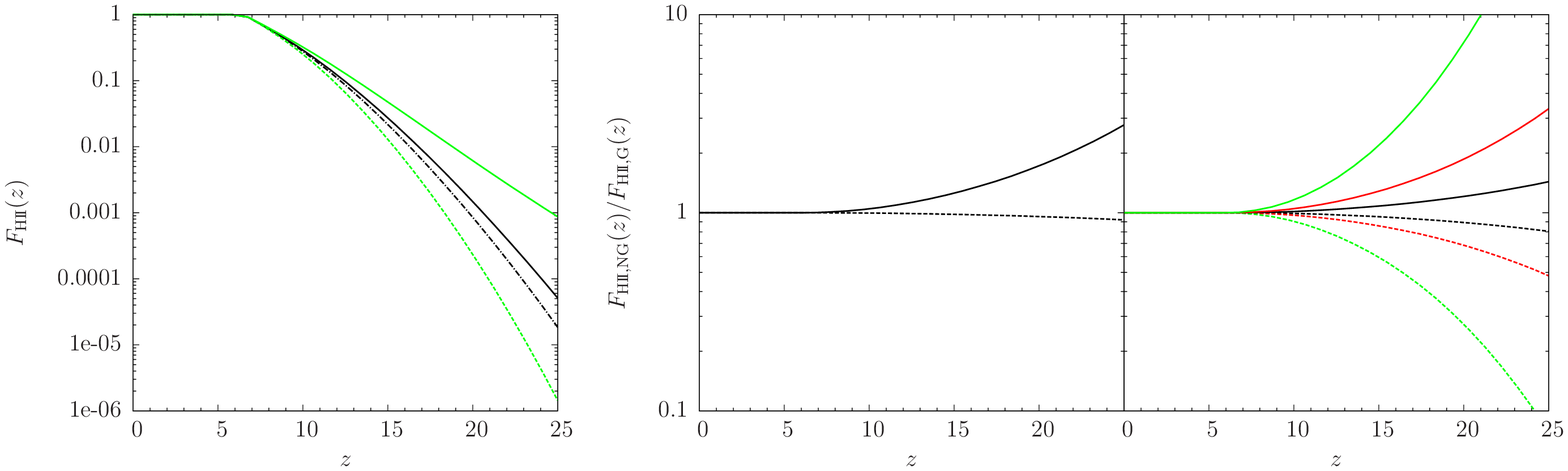}
\caption{ 
The left panel shows the evolution of the ionized fraction $F_{\rm HI\!I}$
for the Gaussian and two different non-Gaussian models here
considered, as in the top panels of Fig. \ref{fig:3a}.  The two panels
on the right show the ratio between the ionized fraction
$F_{\rm HI\!I}$ for non-Gaussian and Gaussian models as a function of
redshift.  In the left panel, the results for non-Gaussian models with
local shape is shown, where $f_{\rm NL}=-9$ (dashed lines) and $f_{\rm
NL}=111$ (solid lines) are assumed. In the right panel, which refers
to non-Gaussian models with equilateral shape, dashed and solid lines
correspond to $f_{\rm NL}=-151,253$, respectively; black, red and
green lines refer to $\alpha=0$ (i.e. no scale-dependence),
$\alpha=-0.1$ and $\alpha=-0.2$, respectively.  }
\label{fig:3b}
\end{figure*}

The evolution of the collapsed fraction due to non-Gaussianity
strongly affects the amount of ionized IGM at different cosmological
epochs.  As illustred in Fig.\ref{fig:3b}, that shows the evolution of
the filling factor both for local and equilateral models, a positive
primordial non-Gaussianity produces, if a scale-dependent $f_{\rm NL}$
parameter is assumed, an increase of the ionized IGM density, that can
become 5 times higher than that predicted in the Gaussian case.  The
IGM reionization results slower when negative values are considered,
since the smaller collapsed fraction produces a mild evolution of the
filling factor, that is smaller compared to the Gaussian case, at
every cosmological epoch.

\begin{figure*}
\includegraphics[height=5cm, width=17cm]{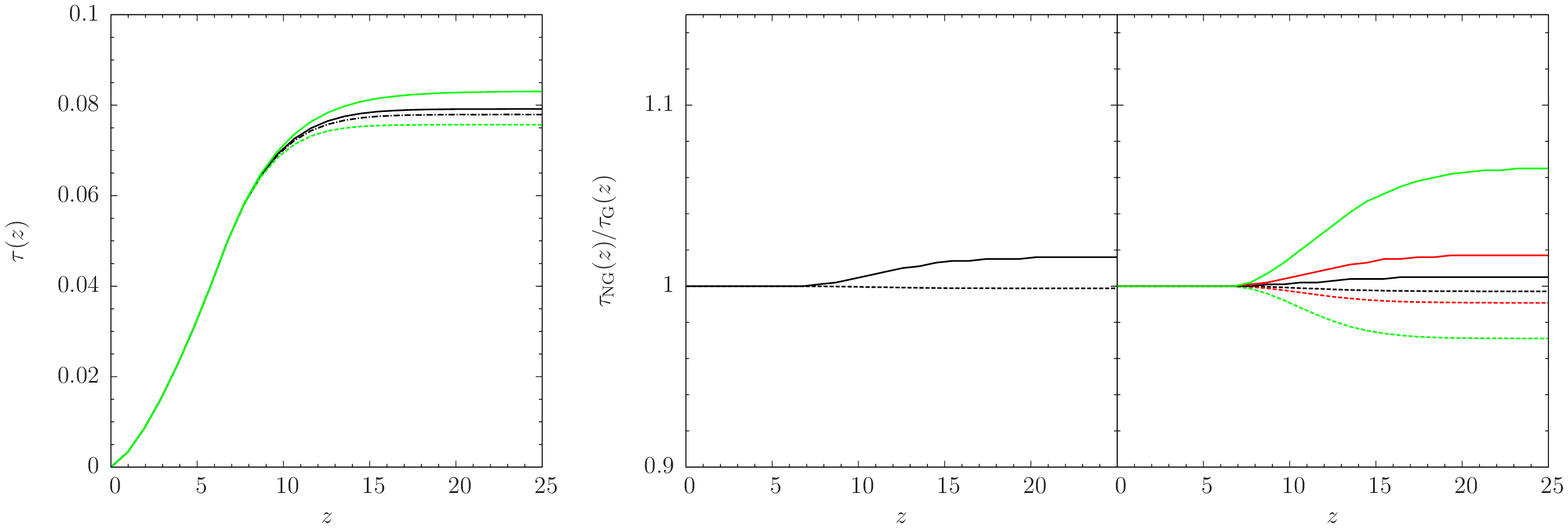}
\caption{
The left panel shows the evolution of the IGM optical depth $\tau$ for
the Gaussian and two different non-Gaussian models here considered, as
in the top panels of Fig. \ref{fig:3a}.  The two panels on the right
show the ratio between the reionization optical depth $\tau$ for
non-Gaussian and Gaussian models as a function of redshift.  In the
left panel the results for non-Gaussian models with local shape is
shown, where $f_{\rm NL}=-9$ (dashed lines) and $f_{\rm NL}=111$
(solid lines) are assumed. In the right panel, which refers to
non-Gaussian models with equilateral shape, dashed and solid lines
correspond to $f_{\rm NL}=-151,253$, respectively; black, red and
green lines refer to $\alpha=0$ (i.e. no scale-dependence),
$\alpha=-0.1$ and $\alpha=-0.2$, respectively.}
\label{fig:4b}
\end{figure*}

Finally in Fig.\ref{fig:4b} we show the effects of the primordial
non-Gaussianity on the reionization optical depth $\tau$ assuming both local
and equilateral models. As a consequence of the evolution of the
ionized fraction, in the initial phases of reionization the values of
$\tau$ for the models with positive $f_{\rm NL}$ are higher than those
predicted for the Gaussian case. In particular, when a
scale-dependent non-Gaussianity is assumed, the change in $\tau$ is
higher than 10 per cent for $\alpha<-0.2$. In this case, the
non-Gaussian optical depth would be $\tau \sim 0.083$ at $z\sim 30$,
that is still consistent with the last WMAP observations
($\tau=0.084\pm0.016$).  Assuming a local shape with largest positive
$f_{\rm NL}$ here considered we find $\tau \sim 0.079$. Notice that
for the Gaussian model we predict a reionization optical depth at $z=30$
$\tau=0.078$.

We notice that our results are in qualitative agreement with the reionisation
picture resulting from the analysis performed by \cite{chen2003}. However, a direct
comparison cannot be done because of the differences in the considered
cosmological parameters: in particular they assume the ones suggested by
the first-year WMAP analysis, with a significantly higher power spectrum normalization ($\sigma_8=0.9$).
Furthermore, they parametrize the degree of non-Gaussianity using the
expectation value of the considered modified Poisson distribution $\lambda$,
that is inversely proportional to the $f_{\rm NL}$ parameter here
used: $\lambda \propto 1/f_{\rm NL}$. 
Their results show that a high amount of primordial non-Gaussianity
(small $\lambda$, i.e. high $f_{\rm NL}$) produces large deviations with
respect to the Gaussian case in the ionized fraction and the IGM optical
depth, in agreement with our results. We should, however, notice that
for the Gaussian reference model the estimate of $\tau$ in \cite{chen2003}
differs by 20 per cent with respect the value we predict using the 5-year WMAP cosmology.

\begin{figure*}
\includegraphics[height=6cm, width=8cm]{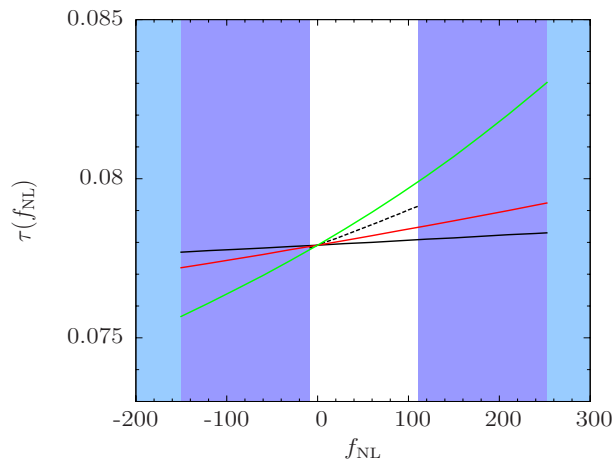}
\caption{ 
The reionization optical depth $\tau$ at redshift $z=30$ as predicted
in different non-Gaussian scenarios is shown as a function of $f_{\rm
NL}$. The dashed black curve refers to the local case, while the solid
black, red and green lines correspond to the equilateral case with
$\alpha=0$, $\alpha=-0.1$ and $\alpha=-0.2$, respectively. The blue
and cyan rectangles represent the areas currently excluded by the WMAP
observations for the local and equilateral cases, respectively.}
\label{fig:6b}
\end{figure*}

The predicted reionization optical depth (at $z=30$) as a function of
the non-Gaussianity parameter $f_{\rm NL}$ is shown in
Fig.\ref{fig:6b} for the local and equilateral cases. Indeed, the
standard deviation on the current estimate of $\tau$ from the last
WMAP results ($\sim 20$ per cent) does not allow us to strictly
constrain the scale dependence of non-Gaussianity and, as shown by
\cite{liguori2008}, should affect the $f_{\rm NL}$ value by a $\sim 3$ per cent
and $\sim 5$ per cent uncertainty for the local and equilateral shapes,
respectively.  We estimate that in order to distinguish among the
different scale dependences, a precision between $\sim 1$ per cent
($f_{\rm NL}<0$) and $\sim 8$ per cent ($f_{\rm NL}>0$) on the $\tau$ measurement is
required, using CMB-like experiments. In this sense, the smaller
standard deviations expected by Planck ($\sim 6$ per cent)
\citep{mukherjee2008} could better probe the reionization optical
depth and would possibly constrain the non-Gaussianity parameter with
an uncertainty of $\sim 1$ per cent and $\sim 2$ per cent in the local and
equilaterals models, respectively. We should however keep in mind that
the current WMAP estimate of $\tau$ cited in this paper is affected by
cosmic variance, since it is based on the bispectrum of the cosmic
microwave background.  Recent works investigate alternative methods to
avoid this problem \citep[see, e.g.][]{seljak2008}.

\begin{figure*}
\includegraphics[height=5cm, width=17cm]{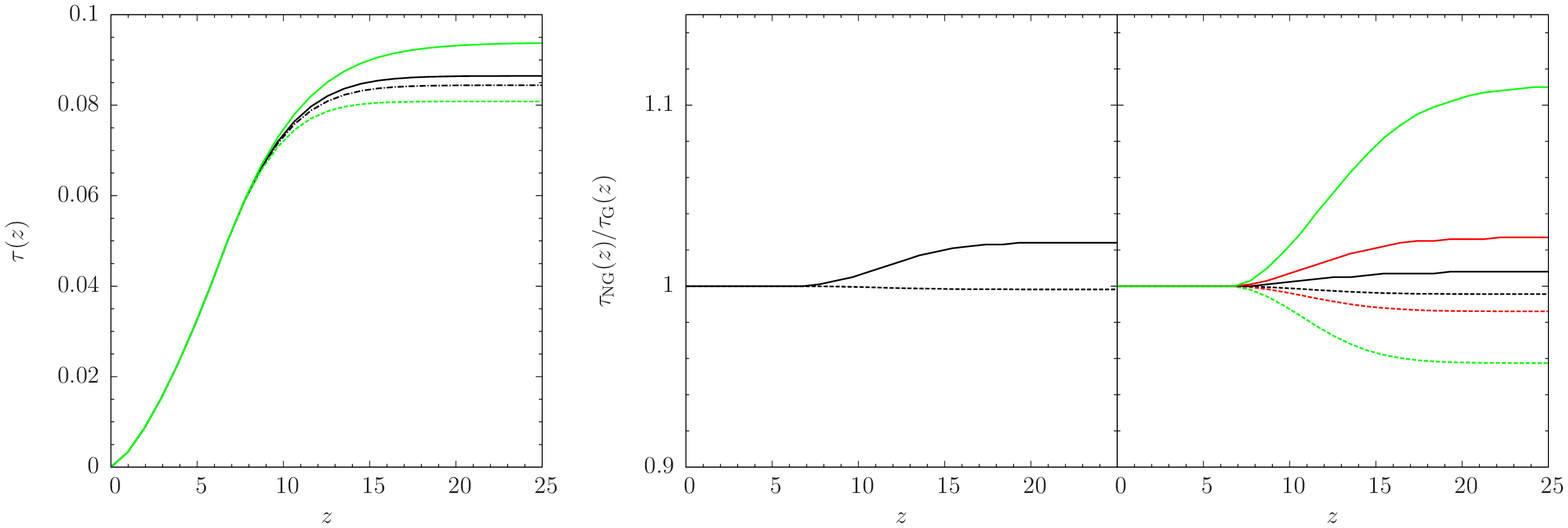}
\caption{
  As in Fig.\ref{fig:4b}, but assuming $\beta=2$ in the redshift
  dependence of the clumping factor (see
  eq.\ref{eq:2c}).}\label{fig:5b}
\end{figure*}

As a general warning, we stress that the description of the
recombination process in the analytic model here adopted relies on
some simplifying assumptions which can affect the way in which
reionization occurs. A deeper investigation of the IGM statistical
distribution would require the analysis of suitable numerical
simulations having non-Gaussian initial conditions.  As an example of
possible biases introduced by the model uncertainties, in
Fig.\ref{fig:5b} we show how a time-evolving clumping factor impacts
the reionization optical depth. The results have been obtained by
setting $\beta=2$ in eq.(\ref{eq:2c}), as suggested by
\cite{avelino2006}, for which the reionization optical depth is in
agreement with the 3-year WMAP results for many degrees of
non-Gaussianity.

  In order to estimate the relative importance of the details of
  the analytic modelling of the reionization process here adopted, we
  can compare the IGM optical depths predicted assuming $\beta=0$
  (Fig.\ref{fig:4b}) and $\beta=2$ (Fig.\ref{fig:5b}).  As shown by
  the corresponding left panels, for the Gaussian model the effect on
  $\tau$ due to the change in $\beta$ and then to the unknown IGM
  physics is approximately of 10 per cent. Viceversa, considering the
  non-Gaussian to Gaussian ratios (shown in the right panels of
  Figs.\ref{fig:4b} and \ref{fig:5b}), changing $\beta$ from 0 to 2
  produces negligible effects on the results. In fact we find that at
  high $z$ the ratios obtained with $\beta=0$ and $\beta=2$ differ by
  less than 1 per cent for the mildly non-Gaussian models and by $\sim
  4$ per cent for the non-Gaussian models with the most extreme
  scale-dependence ($f_{\rm NL}=253$). By looking at the right panel
  of Fig. \ref{fig:5b}, we notice that for this last model the effect
  of introducing a primordial non-Gaussianity can rise up to 10 per
  cent.  This shows that the non-Gaussian to Gaussian ratios are only
  mildly affected by the choice of the recombination model: this
  justifies the fact that in this paper we preferred to show most of
  the results in terms of ratios in spite of absolute values.

Finally, we remark that our approach, which is 
based on the PS74 formalism and its extensions, cannot fully account
for the source clustering, that could indeed have relevant effects
on the morphology of the HII regions. Since the source distribution
depends on the clustering amplitude through the bias parameter, that
is different if a primordial non-Gaussianity is considered \citep[see,
e.g.][]{matarrese2008, dalal2008,carbone2008}, we expect that our
  results could be affected by this approximation, which can be
  improved only with suitable numerical simulations.

%%%%%%%%%%%%%%%%%%%%%%%%%%%%%%%%%%%%%%%%%%%%%%%%%%%%%%%%%%%%%%%%%%%
%%%%%%%%%%%%%%%%%%%%%%%% coonclusion %%%%%%%%%%%%%%%%%%%%%%%%%%%%%%
%%%%%%%%%%%%%%%%%%%%%%%%%%%%%%%%%%%%%%%%%%%%%%%%%%%%%%%%%%%%%%%%%%%

\section{Conclusions} \label{sect:conclu}

The aim of this work was to investigate how primordial
non-Gaussianity may alter the reionization history when compared
to the standard scenario based on Gaussian statistics.  We have chosen a
simple analytic method to describe the physical processes acting on
the IGM to make predictions on the evolution of the ionized fraction
and the reionization optical depth.  Our work extends previous analyses
\citep{chen2003,avelino2006} based on simplified ways to introduce primordial
non-Gaussianity, considering models motivated by inflation. In
particular we assume two different hierarchical evolution scenarios for
the ionizing sources, characterized by scale-independent and
scale-dependent non-Gaussianity.  All scenarios here considered are
not violating the constraints coming from the recent analysis of the
5-year WMAP data.
 
Our main conclusions can be summarized as follows:

\begin{enumerate}
\renewcommand{\theenumi}{(\arabic{enumi})}
\item non-Gaussianity affects the abundance of the dark matter haloes,
  since the formation of high-mass collapsed objects is enhanced
  (reduced) when positive (negative) values for the $f_{\rm NL}$
  parameter are assumed.  This effect is more evident at earlier
  cosmological epochs, exactly when reionization begins.
\item As a consequence, for positive primordial non-Gaussianity, the
  collapsed fraction in type Ia and Ib sources is higher than for the
  Gaussian case at the same epoch, and the difference increases with
  $z$. The opposite result applies when $f_{\rm NL}<0$ is assumed.
\item The IGM filling factor is higher and its evolution is faster
  than in the Gaussian scenario if a positive $f_{\rm NL}$ is
  assumed. This effect is enhanced for a scale-dependent
  non-Gaussianity, that can produce a 5 times higher $F_{\rm HI\!I}$ with
  respect to the Gaussian case, at early cosmological epochs.
  Viceversa the filling factor is smaller and has a mild redshift
  evolution for negative $f_{\rm NL}$.
\item Both local and equilateral non-Gaussianity have a small (less
  than 10 per cent) impact on the reionization optical depth, but the effect is
  enhanced assuming a scale-dependent $f_{\rm NL}$ parameter.
\end{enumerate}

We finally remark that our predictions of the reionization optical
depth in non-Gaussian cosmologies are in agreement with that estimated
by 5-year WMAP analysis within $1\sigma$ error bars and a precision
higher than that of WMAP is required to constrain non-Gaussianity and
its scale dependence. Ideally one would simulate
  reionization in non-Gaussian cosmological models using
  hydrodynamical simulations that incorporate all the relevant
  physical processes in a consistent framework (most importantly
  radiative transfer effects in the IGM).  However, such approach is
  very time consuming due to the large box size and the high
  resolution required to simulate large volumes and, at the same time,
  the physics of the sources of radiation.  For this reason, some
  approximate semi-analytic schemes such as the one presentend here
  are still useful especially when calibrated on the more robust
  results of the hydrodynamical runs.

%%%%%%%%%%%%%%%%%%%%%%%%%%%%%%%%%%%%%%%%%%%%%%%%%%%%%%%%%%%%%%%%%%%%%%%%
%%%%%%%%%%%acknowledgements%%%%%%%%%%%%%%%%%%%%%%%%%%%%%%%%%%%%%%%%%%%%%
%%%%%%%%%%%%%%%%%%%%%%%%%%%%%%%%%%%%%%%%%%%%%%%%%%%%%%%%%%%%%%%%%%%%%%%%
\section*{acknowledgements}

We acknowledge financial contribution from contracts ASI-INAF
I/023/05/0, ASI-INAF I/088/06/0 and ASI-INAF I/016/07/0.  We thank
Adam Lidz and Enzo Branchini for useful discussions, and the
  anonymous referee for his/her comments which help us to improve the
  presentation of our results.

%%%%%%%%%%%%%%%%%%%%%%%%%%%%%%%%%%%%%%%%%%%%%%%%%%%%%%%%%%%%%%%%%%%%%%%%
%%%%%%%%%%%%%%%%%%%%%%%%%%bibliography%%%%%%%%%%%%%%%%%%%%%%%%%%%%%%%%%%
%%%%%%%%%%%%%%%%%%%%%%%%%%%%%%%%%%%%%%%%%%%%%%%%%%%%%%%%%%%%%%%%%%%%%%%%
%\bibliographystyle{mn2e}
%\bibliography{master2.bib}
\newcommand{\noopsort}[1]{}

\label{lastpage}
\end{document}